\documentstyle[aps,preprint,prb]{revtex}
\begin{document}
\def\vk{\vec k} 
\def\br{{\bf r}}
\title{\bf Strong-Coupling Theory of Impure Superconductors : Correspondence
with Weak-Coupling Theory}
\author{Yong-Jihn Kim }
\address{Department of Physics,  Korea Advanced Institute of Science and 
Technology,\\
 Taejon 305-701, Korea}
\maketitle
\begin{abstract}
We reconsider the ordinary impurity effect on the transition temperature 
$T_{c}$ of superconductors using the Eliashberg formalism.
It is shown that the correspondence principle, which relates strong-coupling 
and weak-coupling theories, works only when Anderson's pairing condition 
between the time-reversed scattered-states is used.
For an Einstein phonon model, the change of the electron density
of states caused by the impurity scattering leads to a $T_{c}$ decrease
 proportional to ${1/ E_{F}\tau}$ in the dirty limit.  
It is pointed out that the phonon-mediated interaction decreases 
by the same weak localization correction term as that of the conductivity. 
Accordingly, for strongly localized states the phonon-mediated interaction 
is exponentially small.
We also discuss the case of Debye phonon model.

\vspace{1pc}

\end{abstract}

\vfill\eject
\section{\bf Introduction} 

Recently Kim and Overhauser (KO)$^{1}$ showed the following: 

(i) Abrikosov and Gor'kov's (AG) theory$^{2}$ of an impure superconductor 
predicts a large decrease of $T_{c}$, proportional to $1/\omega_{D}\tau$. 
$\omega_{D}$ denotes the Debye frequency
and $\tau$ is the scattering time, respectively.

(ii) Anderson's theorem$^{3}$ is valid only to the first power in the impurity
concentration. For strongly localized states, the phonon-mediated interaction
is exponentially small.

\noindent
The existence of the above correction term was confirmed by Abrikosov, Gor'kov
and Dzyaloshinskii,$^{4}$  and was also shown by other workers.$^{5-7}$ 
The correction term is related with the change of electron density of states
due to the impurity scattering.
However, the correct value was shown to be $1/E_{F}\tau$.$^{1}$
Here $E_{F}$ denotes the Fermi energy.
In their comment on Ref. 1, Abrikosov and Gor'kov$^{8}$ argued that  the
correction term disappears in the Eliashberg equation apart from the
corrections of the order  $1/E_{F}\tau$. 
In fact, this result was first obtained by Tsuneto.$^{9}$
As a result, they admittedly showed that Gor'kov formalism is inconsistent
with the Eliashberg equation.

At this point, we may need to pause to answer the following deep 
question:
Is there a correspondence rule between strong-coupling and weak-coupling theories of impure superconductors?
The answer is yes.
It is well-known that the correspondence principle was very helpful in 
developing quantum mechanics.
The purpose of this paper is to show that the correspondence principle, which
relates strong-coupling and weak-coupling theories, works provided that 
Anderson's pairing condition is satisfied.
Then, the result of strong-coupling theory with an Einstein phonon model 
leads to that of weak-coupling theory  in the static limit. 

In this study, because we disregard the change of phonon spectrum and Coulomb 
interaction due to the impurities,
impurity scattering can affect the $T_{c}$ of superconductors 
only by changing the electron density of states $N_{o}$ and the phonon-mediated interaction $V$.
For an Einstein phonon model and in the dirty limit, we show that the change 
of electron density of states caused by the impurity scattering leads to 
a $T_{c}$ decrease proportional to ${1/ E_{F}\tau}$.  
When weak localization becomes important, the 
phonon-mediated interaction decreases by the same correction term as that 
of the conductivity. 
Accordingly, for strongly localized states the phonon-mediated interaction 
is exponentially small.

The failure of AG theory comes from the inadequate treatment of pairing 
constraint on the self-consistency equation of Gor'kov formalism.$^{11}$
Although both Gor'kov formalism and the Bogoliubov-de Gennes equations 
pair the electrons in states which are linear combinations of the normal states,$^{11,12}$
the physical constraint of the Anomalous Green's function
leads to Anderson's pairing condition.$^{11}$

\section{\bf Strong-coupling theory with Anderson's pairing} 

We follow the real space formalism of the strong-coupling theory
by Eilenberger and Ambegaokar.$^{13}$ 
(See also refs. 14-18.) 
The Hamiltonian for the electron-phonon interaction takes the form
\begin{equation}
H_{int} = \gamma \int \psi^{+}({\bf r})\psi ({\bf r})\phi({\bf r})d{\bf r}, 
\end{equation}
where $\Psi({\bf r})$ and $\phi({\bf r})$ are the electron and phonon field 
operators. $\gamma$ is the coupling constant.
The equations of the motion for the thermodynamic Green's functions
$G(\omega_{n}, {\bf r}, {\bf r'})$ and $F^{+}(\omega_{n}, {\bf r}, {\bf r'})$
are given
\begin{eqnarray}
(i\omega_{n} &+& {1\over 2m}\nabla^{2} + V({\bf r}) + \mu)
G(\omega_{n}, {\bf r}, {\bf r'}) \nonumber\\ 
&=& \delta({\bf r}-{\bf r'}) + \gamma^{2}\int 
d{\bf r}_{o}
\Sigma(\omega_{n}, {\bf r},{\bf r}_{o}) G(\omega_{n}, {\bf r}_{o}, {\bf r'})
\nonumber \\
& & + \gamma^{2}\int d{\bf r}_{o} \phi(\omega_{n}, {\bf r},{\bf r}_{o})
F^{+}(\omega_{n}, {\bf r}_{o}, {\bf r'}),
\end{eqnarray}
\begin{eqnarray}
(-i\omega_{n} &+& {1\over 2m}\nabla^{2} + V({\bf r}) + \mu)
F^{+}(\omega_{n}, {\bf r}, {\bf r'})  \nonumber\\
&=&  \gamma^{2}\int d{\bf r}_{o}
\Sigma^{+}(\omega_{n}, {\bf r},{\bf r}_{o})
F^{+}(\omega_{n}, {\bf r}_{o}, {\bf r'})\nonumber\\
& &+ \gamma^{2}\int d{\bf r}_{o}
\phi^{+}(\omega_{n}, {\bf r},{\bf r}_{o})
G(\omega_{n}, {\bf r}_{o}, {\bf r'}),
\end{eqnarray}
where 
\begin{equation}
 \Sigma(\omega_{n}, {\bf r},{\bf r'}) = T\sum_{n'} D(\omega_{n}, \omega_{n'},
{\bf r}, {\bf r'})G(\omega_{n'}, {\bf r}, {\bf r'}), 
\end{equation}
\begin{equation}
 \phi(\omega_{n}, {\bf r},{\bf r'}) = T\sum_{n'} D(\omega_{n}, \omega_{n'},
{\bf r}, {\bf r'})F(\omega_{n'}, {\bf r}, {\bf r'}). 
\end{equation}
$V({\bf r})=\sum_{i}V_{o}\delta({\bf r}-{\bf R}_{i})$ is the scattering 
potential of the impurities and
$\omega_{n} = (2n+1)\pi T$. $D$ is the phonon Green's function.

It is usually assumed that the electron-phonon interaction is 
local,$^{20, 13, 14}$
(i.e.),
\begin{equation}
D(\omega_{n},\omega_{n'}, {\bf r},{\bf r'}) = \delta({\bf r}-{\bf r'})
\lambda(\omega_{n},\omega_{n'}),
\end{equation}
\begin{equation}
 \Sigma(\omega_{n}, {\bf r},{\bf r'}) = 
 \delta({\bf r}-{\bf r'}) \Sigma(\omega_{n}, {\bf r}), 
\end{equation}
\begin{equation}
 \phi(\omega_{n}, {\bf r},{\bf r'}) = 
 \delta({\bf r}-{\bf r'}) 
 \phi(\omega_{n}, {\bf r}). 
\end{equation}
This approximation is exact for an Einstein phonon model.
Then $ \lambda(\omega_{n},\omega_{n'})$ is given by
\begin{equation}
\lambda(\omega_{n},\omega_{n'})={\omega_{D}^{2}\over \omega_{D}^{2}+
(\omega_{n}-\omega_{n'})^{2}}.
\end{equation}
For Debye phonon model the pairing interaction is nonlocal,
which is discussed below.

The normal-state Green's function $G_{N} (= G_{N}^{\uparrow}=G_{N}^{\downarrow})$
satisfies the equation
\begin{equation}
(i\omega_{n} + {1\over 2m}\nabla^{2} + V({\bf r}) + \mu)
G_{N}(\omega_{n}, {\bf r}, {\bf r'})  = \delta({\bf r}-{\bf r'}) + \gamma^{2}\int 
d{\bf r}_{o} \Sigma(\omega_{n}, {\bf r},{\bf r}_{o})
G_{N}(\omega_{n}, {\bf r}_{o}, {\bf r'}) ,
\end{equation}
and it is given by
\begin{equation}
G_{N}(\omega_{n}, {\bf r},{\bf r'}) = \sum_{m}{\psi_{m}({\bf r})\psi^{*}_{m}
({\bf r'})\over i\omega_{n}Z(\omega_{n})-\epsilon_{m}}, 
\end{equation}
where $Z(\omega_{n})$ is the renormalization factor and $\psi_{m}({\bf r})$ is 
the scattered eigenstate.
From Eqs. (3) and (10) the Anomalous Green's 
function $F^{+}(\omega_{n}, {\bf r},{\bf r'})$, near the transition temperature,
 can be rewritten in the form
\begin{equation}
F^{+}(\omega_{n}, {\bf r},{\bf r'}) = \gamma^{2}\int d{\bf r}_{o}G_{N}^{\uparrow}(-\omega_{n},
{\bf r}_{o},{\bf r})\phi^{+}(\omega_{n},{\bf r}_{o})
G_{N}^{\downarrow}(\omega_{n},{\bf r}_{o},{\bf r'}).
\end{equation}
Accordingly, we obtain the self-consistency equation for $\phi^{+}$
\begin{eqnarray}
\phi^{+}(\omega_{n}, {\bf r}) & = & T\sum_{n'}\lambda
(\omega_{n},\omega_{n'}) F^{+}(\omega_{n'}, {\bf r},{\bf r})\nonumber\\
& = & \gamma^{2}T\sum_{n'}\lambda(\omega_{n},\omega_{n'})\int d{\bf r}_{o}
G_{N}^{\uparrow}(-\omega_{n'},{\bf r}_{o},
{\bf r})G_{N}^{\downarrow}(\omega_{n'},{\bf r}_{o},{\bf r})\phi^{+}(\omega_{n'},{\bf r}_{o}).
\end{eqnarray}

The pair potential $\Delta^{*}(\omega_{n},{\bf r})$ is defined by
 $\Delta^{*}(\omega_{n},{\bf r}) = \phi^{+}(\omega_{n},{\bf r}) / Z(\omega_{n})$.
Therefore we find the self-consistency equation for the pair potential
to be$^{13}$
\begin{eqnarray}
\Delta^{*}(&\omega_{n}&, {\bf r})Z(\omega_{n}) \nonumber\\
 &=& \gamma^{2}  T\sum_{n'}\lambda(\omega_{n},\omega_{n'})\int d{\bf r}_{o}
G_{N}^{\uparrow}(-\omega_{n'},{\bf r}_{o},
{\bf r})G_{N}^{\downarrow}(\omega_{n'},{\bf r}_{o},{\bf r})\Delta^{*}(\omega_{n'},{\bf r}_{o})
Z(\omega_{n'}).  
\end{eqnarray}
Equation (14) states physically that the pair potential $\Delta^{*}(\omega_{n'},{\bf r}_{o})$
launches (from the regions near ${\bf r}_{o}$) electron pairs which collaborate to generate
a pair potential $\Delta^{*}(\omega_{n},{\bf r})$ in the region near ${\bf r}$.
However, Eq. (14) misses the most important information of Anderson's pairing condition.
If we substitute Eq. (11) into Eq. (14), we find extra pairings between 
$m\uparrow$ and $m'(\not= \overline{m})\downarrow$.
$\overline{m}$ denotes the time reversed partner of the scattered state
$m$.
Whereas it was shown$^{19}$ that Anderson's pairing condition is 
derived from the physical constraint of the Anomalous Green's function, i.e.,
\begin{eqnarray}
\overline{F^{+}(\omega_{n},{\bf r},{\bf r'})}^{imp} &=& 
\overline{F^{+}(\omega_{n},{\bf r}-{\bf r'})}^{imp}, \\
\overline{\Delta^{*}(\omega_{n},{\bf r})}^{imp} &=& 
\overline{ \Delta^{*}(\omega_{n})}^{imp}.
\end{eqnarray}
$({\bar{\ \ } }^{imp})$ means an average over impurity positions.
Consequently, the revised self-consistency equation is
\begin{eqnarray}
\Delta^{*}(&\omega_{n}&, {\bf r})Z(\omega_{n})  = \nonumber\\
 & \gamma^{2} & T\sum_{n'}\lambda(\omega_{n},\omega_{n'})\int d{\bf r}_{o}
\{G_{N}^{\uparrow}(-\omega_{n'},{\bf r}_{o}, {\bf r})
G_{N}^{\downarrow}(\omega_{n'},{\bf r}_{o},{\bf r})\}^{P}
\Delta^{*}(\omega_{n'},{\bf r}_{o})Z(\omega_{n'}), 
\end{eqnarray}
where
$P$ denotes Anderson's pairing constraint. 

The importance of Anderson's pairing constraint was already noticed by Ma and Lee.$^{21}$
They showed that the gap parameter is given by
\begin{equation}
\Delta^{*}(\omega_{n}, m) = \int \psi_{m}({\bf r})\psi^{*}_{m}({\bf r})
\Delta^{*}(\omega_{n},{\bf r})d{\bf r}.
\end{equation}
Substitution of Eq. (17) into Eq. (18) leads to a strong-coupling gap equation
\begin{equation}
\Delta^{*}(\omega_{n}, m)Z(\omega_{n}) = T
\sum_{n'}\lambda(\omega_{n},\omega_{n'})
 \sum_{m'}V_{mm'}{\Delta^{*}(\omega_{n'},m')Z(\omega_{n'})\over [\omega_{n'}Z(\omega_{n'})]^{2}
+\epsilon_{m'}^{2}},
\end{equation}
where
\begin{equation}
V_{mm'} = \gamma^{2}\int |\psi_{m}({\bf r})|^{2}
 |\psi_{m'}({\bf r})|^{2}d{\bf r}.
\end{equation}
$\epsilon_{m'}$ denotes the eigenenergy.

\section{\bf Weak-coupling limit} 

The strong-coupling theory leads to the weak-coupling theory in the static limit, (i.e.),
\begin{eqnarray}
\Delta^{*}(\omega_{n},{\bf r})&=&\Delta^{*}(0, {\bf r})=\Delta^{*}({\bf r}),\\
Z(\omega)&=&Z(0)=1,\\
\lambda(\omega_{n},\omega_{n'})&=&\lambda(0,0)=1.
\end{eqnarray}
In BCS theory, the retardation effect is taken into account by a cutoff of the matrix
element.$^{3}$ 
Anderson emphasized that the attractive region is a function not of $\epsilon_{\vec k}$,
the energy of the plane wave states, but of $\epsilon_{n}$, the energy of scattered
states.$^{3}$
It was also shown that Gor'kov formalism should use the BCS cutoff in the eigenenergies 
not in the momentum state energies.$^{1,23}$

From Eqs. (14), (17), and (19), we find
\begin{equation}
\Delta^{*}( {\bf r})= \gamma^{2}  T\sum_{n'}\int d{\bf r}_{o}
G_{N}^{\uparrow}(-\omega_{n'},{\bf r}_{o},
{\bf r})G_{N}^{\downarrow}(\omega_{n'},{\bf r}_{o},{\bf r})\Delta^{*}({\bf r}_{o}) ,  
\end{equation}
\begin{equation}
\Delta^{*}({\bf r}) =  \gamma^{2} T\sum_{n'}\int d{\bf r}_{o}
\{G_{N}^{\uparrow}(-\omega_{n'},{\bf r}_{o}, {\bf r})
G_{N}^{\downarrow}(\omega_{n'},{\bf r}_{o},{\bf r})\}^{P}
\Delta^{*}({\bf r}_{o}), 
\end{equation}
and

\begin{equation}
\Delta^{*}(m) = T
\sum_{n'} \sum_{m'}V_{mm'}{\Delta^{*}(m')\over \omega_{n'}^{2}
+\epsilon_{m'}^{2}}.
\end{equation}

Equation (26) may be rewritten in the familiar form
\begin{equation}
\Delta^{*}(m) = \sum_{m'}V_{mm'}{\Delta^{*}(m') \over 2\epsilon_{m'}} 
tanh({\epsilon_{m'}\over 2T}),
\end{equation}
since$^{22}$ 
\begin{equation}
T\sum_{n'}{1\over \omega_{n'}^{2}+\epsilon^{2}} = {1\over 2\epsilon}tanh{\epsilon\over 2T}.
\end{equation}
Note that Eq. (27) is the linearized BCS gap equation near $T_{c}$.

\section{\bf Theory of Impure superconductors} 

Now we discuss the impurity effect on the transition temperature of 
superconductors.
For a pure system, the coupling constants are given by
\begin{eqnarray}
\lambda &=& N_{o}V, \hspace{8pc} {\rm \ BCS\ theory} \\
\lambda&=& N_{o}\gamma^{2}=N_{o}{g^{2}\over M\omega_{D}^{2}},
 \hspace{3pc} {\rm strong-coupling\ theory}
\end{eqnarray}
where $g^{2}$ is the average square electronic matrix 
element in McMillan's notation,$^{10}$ and $M$ is the ionic mass. 
Because we disregard the change of phonon spectrum and Coulomb 
interaction due to the impurities,
impurity scattering can affect the $T_{c}$ of superconductors 
only by changing $N_{o}$ and/or $V$ (or $g^{2}$).
The coupling constants for impure superconductors 
lead to
\begin{eqnarray}
\lambda &=& N'_{o}<V_{mm'}>, \hspace{12pc} {\rm \ \ BCS\ theory} \\
\lambda &=& N'_{o}{g^{2}\over M\omega_{D}^{2}}<\int |\psi_{m}({\bf r})|^{2}
 |\psi_{m'}({\bf r})|^{2}d{\bf r}>, \hspace{3pc} {\rm strong-coupling\ theory}
\end{eqnarray}
where $N_{o}'$ is the density of states for impure superconductors.
The angular brackets indicate an impurity average.
As will be shown below, $\lambda$ does not depend on the energies. 
Notice that the coupling constants 
are the basically same both in weak and
strong-coupling theories. 
Accordingly, both the strong-coupling gap equation (19) 
and the BCS gap qeaution (27) give the basically same result.
It dose not matter whether the retardation effect is taken into account by the
phonon Green's function $\lambda (\omega_{n},\omega_{n'})$ or
the BCS cutoff of the matrix elements contrary to AG's recent claim.$^{8}$

Eqs. (31) and (32) show that the most important quantity is $V_{mm'}$ 
which determines the change of the electron-phonon interaction due to the impurities.
In the dirty limit where the mean free path $\ell$ is $\sim 100\AA$, Anderson's theorem 
is valid, (i.e.),  
\begin{equation}
V_{mm'} = \gamma^{2} = {g^{2}\over M\omega^{2}_{D}}= V.
\end{equation}
Therefore, $T_{c}$ does not change due to the impurities.
On the other hand, Kim and Overhauser$^{1}$ showed that $V_{mm'}$ is 
exponentially small for the
strongly localized states.$^{1}$
It is, then, expected that $V_{mm'}$ decreases by weak localization.
In fact, the same weak localization correction terms occur both in the
conductivity and the phonon-mediated interaction.$^{11,12}$
Table I shows $\ell$ and $V_{mm'}$ for different disorder limits.

For thin films, the empirical formula is given$^{24}$
\begin{equation}
{T_{co}-T_{c}\over T_{co}} \propto {1\over k_{F}\ell} \propto R_{\Box},
\end{equation}
where $T_{co}$ is the unperturbed value of $T_{c}$ and $R_{\Box}$ is the sheet resistance.
On the other hand, bulk materials show$^{25,26}$ 
\begin{equation}
{T_{co}-T_{c}\over T_{co}} \propto {1\over (k_{F}\ell)^{2}}.
\end{equation}
Notice that these results are obtained if we substitute the matrix elements in Table I 
into the (strong-coupling or weak-coupling) gap equation.
More details will be published elsewhere.$^{27}$

Scattering of conduction electrons by the impurities leads to
a decrease in the electron density of states $N_{o}$ at the Fermi level.
However, this effect is small. The reduced density of states was shown to
be$^{1}$
\begin{equation}
N_{o}'\cong N_{o}(1-{\hbar\over \pi E_{F}\tau}).
\end{equation}
Then, both strong-coupling and weak-coupling gap equations give rise to
\begin{equation}
T_{c}\cong T_{co}- T_{co}{1\over \lambda}{\hbar\over \pi E_{F}\tau}.
\end{equation}
The correction term is negligible, since
\begin{equation}
{\hbar\over E_{F}\tau}<10^{-2},
\end{equation}
for a $1\%$ typical solute.
In the weak localization limit, this correction term may not be small.
However, $T_{c}$ reduction versus $1/E_{F}\tau$ is quadratic not linear for
bulk materials. It seems that the change of the density of states may saturate 
before the weak localization limit is reached.
Nevertheless the above correction term may be important for materials with very
narrow bands.$^{28,29}$

\section{Previous Approaches}

The previous approaches used the conventional strong-coupling and weak-coupling self-consistency
equations (14) and (24).$^{2,4-9}$ Accordingly, the previous approaches do not use Anderson's pairing
but pair the electrons in states which are linear combinations of the scattered states.$^{11,12}$
Then $T_{c}$ does not change even if the scattered states are localized. Note that
the linear combination of localized states becomes extended one.
A similar problem was found in Gor'kov and Galitski's (GG)$^{30}$ solution for the d-wave
BCS theory. Using the Gor'kov's formalism without pairing constraint, GG obtained a solution
which is a superposition of several distinct types of the off-diagonal-long-range-order.$^{31}$ 
Their solution was proven to be wrong.$^{31-33}$

From Gor'kov's self-consistency equation (24), Abrikosov and Gor'kov (AG)$^{2}$ showed
\begin{equation}
\overline{\Delta^{*}( {\bf r})}^{imp}= \gamma^{2}  T\sum_{n'}\int d{\bf r}_{o}
\overline{G_{N}^{\uparrow}(-\omega_{n'},{\bf r}_{o},
{\bf r})G_{N}^{\downarrow}(\omega_{n'},{\bf r}_{o},{\bf r})}^{imp}
\overline{\Delta^{*}({\bf r}_{o})}^{imp} ,  
\end{equation}
and
\begin{equation}
1={\gamma^{2}T_{c}\over 8\pi^{3}}\sum_{n'}\int {\eta_{1}\over \omega_{n'}^{2}\eta^{2}_{1}+
\epsilon^{2}}d^{3}k,
\end{equation}
where
\begin{equation}
\eta_{1}=1 + {1\over 2|\omega_{n'}|\tau}.
\end{equation}
The $T_{c}$  decrease is given$^{1,4}$
\begin{equation}
T_{c}\cong T_{co}- T_{co}{1\over \lambda}{\hbar\over \pi \omega_{D}\tau}.
\end{equation}
This result should be compared with the correct result Eq. (37).
AG theory has two problems. One is not using Anderson's pairing and the other is 
using a Dyson equation to find Green's function (with a BCS retardation cutoff)
in the presence of the impurities.
If we use Anderson's pairing, the second problem does not occur. 
In other words, pairing condition is more important.

Tsuneto was the first who considered the strong-coupling theory of impure 
superconductors.$^{9}$
His result may be obtained from Eq. (14) (with $Z=1$), (i.e.),
\begin{eqnarray}
& \overline{\Delta^{*}(\omega_{n}, {\bf r})}^{imp}&\nonumber\\
 & =  \gamma^{2}  T\sum_{n'}& \lambda(\omega_{n},\omega_{n'})\int d{\bf r}_{o}
\overline{G_{N}^{\uparrow}(-\omega_{n'},{\bf r}_{o}, {\bf r})G_{N}^{\downarrow}(\omega_{n'},{\bf r}_{o},{\bf r})}^{imp}
\overline{\Delta^{*}(\omega_{n'},{\bf r}_{o})}^{imp},
\end{eqnarray}
and
\begin{equation}
 \overline{\Delta^{*}(\omega_{n})}^{imp} = {\gamma^{2}T_{c}\over 8\pi^{3}} 
T\sum_{n'}\lambda(\omega_{n},\omega_{n'})
\int {\eta_{1}\over \omega_{n'}^{2}\eta^{2}_{1}+ \epsilon^{2}}
 \overline{\Delta^{*}(\omega_{n'})}^{imp} d^{3}k.
\end{equation}
Equation (44) is very interesting. If we solve the equation, we find
\begin{equation}
T_{c}\cong T_{co}- T_{co}{1\over \lambda}{\hbar\over \pi E_{F}\tau}.
\end{equation}
Whereas the weak-coupling limit of this equation leads to AG's result, 
\begin{equation}
T_{c}\cong T_{co}- T_{co}{1\over \lambda}{\hbar\over \pi \omega_{D}\tau}.
\end{equation}
Consequently, there is no correspondence between weak-coupling and strong-coupling theories. 
Something must be wrong. 

The correct strong-coupling theory needs Anderson's pairing.$^{19}$
From the revised strong-coupling self-consistency and gap equations (17) and (19),
it is given$^{19}$ 
\begin{equation}
\overline{\Delta^{*}(\omega_{n})}^{imp} = T
\sum_{n'}\lambda(\omega_{n},\omega_{n'})
 \sum_{m'}<V_{mm'}>{\overline{\Delta^{*}(\omega_{n'})}^{imp}\over \omega_{n'}^{2}
+\epsilon_{m'}^{2}},
\end{equation}
where
\begin{equation}
<V_{mm'}> = \gamma^{2}<\int |\psi_{m}({\bf r})|^{2}
 |\psi_{m'}({\bf r})|^{2}d{\bf r}>.
\end{equation}
Comparing Eqs. (44) and (47), we find that Tsuneto's result misses
the most important factor $V_{mm'}$, which gives the change of
phonon-mediated interaction due to impurities.
In the weak-coupling limit, one finds
\begin{equation}
1 = T_{c} \sum_{n'}
 \sum_{m'}<V_{mm'}>{1\over \omega_{n'}^{2}
+\epsilon_{m'}^{2}}.
\end{equation}
 In the dirty limit, both Eqs. (47) and (49) lead to
\begin{equation}
T_{c}\cong T_{co}- T_{co}{1\over \lambda}{\hbar\over \pi E_{F}\tau}.
\end{equation}
The correspondence principle is recovered.

\section{Case of Debye Phonon Model}

Now we discuss briefly the strong-coupling theory with Debye phonon model.
Because the pairing interaction is nonlocal, the local approximations
Eqs. (6), (7), and (8) are not valid.
From Eqs. (3) and (10), it is given
\begin{equation}
F^{+}(\omega_{n}, {\bf r},{\bf r'}) = \gamma^{2}\int\int d{\bf r}_{o}
d{\bf r}_{1} G_{N}^{\uparrow}(-\omega_{n},
{\bf r}_{1},{\bf r})\phi^{+}(\omega_{n},{\bf r}_{1},{\bf r}_{o})
G_{N}^{\downarrow}(\omega_{n},{\bf r}_{o},{\bf r'}).
\end{equation}
The self-consistency equation for $\phi^{+}$ leads to
\begin{eqnarray}
\phi^{+}(\omega_{n}, {\bf r}, {\bf r}') & = & T\sum_{n'}
D(\omega_{n},\omega_{n'}, {\bf r}, {\bf r}') 
F^{+}(\omega_{n'}, {\bf r},{\bf r'})\nonumber\\
& = & \gamma^{2}T\sum_{n'}D(\omega_{n},\omega_{n'},{\bf r},{\bf r}')
\int\int d{\bf r}_{o} d{\bf r}_{1}
G_{N}^{\uparrow}(-\omega_{n'},{\bf r}_{1}, {\bf r})
G_{N}^{\downarrow}(\omega_{n'},{\bf r}_{o},{\bf r}')\nonumber\\
&\  & \hspace{2.5in}  \times \ \phi^{+}(\omega_{n'},{\bf r}_{1},{\bf r}_{o}).
\end{eqnarray}
Then the self-consistency equation for the pair potential 
$\Delta^{*}(\omega_{n},{\bf r},{\bf r}')=\phi^{+}(\omega_{n},{\bf r},{\bf r}')/ 
Z(\omega_{n})$ is given 
\begin{eqnarray}
&\Delta^{*}&(\omega_{n}, {\bf r},{\bf r}')Z(\omega_{n}) \nonumber\\
& = & \gamma^{2}T\sum_{n'}D(\omega_{n},\omega_{n'},{\bf r},{\bf r}')
\int\int d{\bf r}_{o} d{\bf r}_{1}
G_{N}^{\uparrow}(-\omega_{n'},{\bf r}_{1}, {\bf r})
G_{N}^{\downarrow}(\omega_{n'},{\bf r}_{o},{\bf r}')\nonumber\\
& \ & \hspace{2.4in}\times \ \Delta^{*}(\omega_{n'},{\bf r}_{1},{\bf r}_{o})
Z(\omega_{n'}).
\end{eqnarray}

Accordingly, the revised self-consistency equation with Anderson's pairing is
\begin{eqnarray}
&\Delta^{*}&(\omega_{n}, {\bf r},{\bf r}')Z(\omega_{n}) \nonumber\\
& = & \gamma^{2}T\sum_{n'}D(\omega_{n},\omega_{n'},{\bf r},{\bf r}')
\int\int d{\bf r}_{o} d{\bf r}_{1}
\{G_{N}^{\uparrow}(-\omega_{n'},{\bf r}_{1}, {\bf r})
G_{N}^{\downarrow}(\omega_{n'},{\bf r}_{o},{\bf r}')\}^{P}\nonumber\\
&\ & \hspace{2.4in} \times \ \Delta^{*}(\omega_{n'},{\bf r}_{1},{\bf r}_{o})
Z(\omega_{n'}).
\end{eqnarray}
Because the gap parameter is given by
\begin{equation}
\Delta^{*}(\omega_{n}, m) = \int\int \psi_{m}({\bf r})\psi_{\overline m}({\bf r}')
\Delta^{*}(\omega_{n},{\bf r},{\bf r}')d{\bf r}d{\bf r}',
\end{equation}
we find a gap equation
\begin{equation}
\Delta^{*}(\omega_{n}, m)Z(\omega_{n}) = T \sum_{n'}\sum_{\vec q}
\lambda(\omega_{n}-\omega_{n'},{\vec q})\sum_{m'}V_{mm',{\vec q}}
{\Delta^{*}(\omega_{n'},m')Z(\omega_{n'})\over [\omega_{n'}Z(\omega_{n'})]^{2}
+\epsilon_{m'}^{2}},
\end{equation}
where
\begin{equation}
\lambda(\omega_{n}-\omega_{n'},{\vec q})=
{\omega_{\vec q}^{2}\over (\omega_{n}-\omega_{n'})^{2} + \omega_{\vec q}^{2}}
\theta (\omega_{D}-\omega_{\vec q}),
\end{equation}
\begin{equation}
V_{mm',{\vec q}} = \gamma^{2}\int\int 
e^{i{\vec q}\cdot ({\bf r}-{\bf r}')}\psi_{m}^{*}({\bf r})\psi^{*}_{\overline m}({\bf r}')
\psi_{m'}({\bf r}) \psi_{\overline m'}({\bf r}')
d{\bf r}d{\bf r}'.
\end{equation}
$\theta$ denotes the usual step function.

In the weak-coupling limit, the revised self-consistency and gap
equations lead to 
\begin{equation}
\Delta^{*}({\bf r},{\bf r}')
 =  \gamma^{2}TV({\bf r}-{\bf r}')\sum_{n'}
\int\int d{\bf r}_{o} d{\bf r}_{1}
\{G_{N}^{\uparrow}(-\omega_{n'},{\bf r}_{1}, {\bf r})
G_{N}^{\downarrow}(\omega_{n'},{\bf r}_{o},{\bf r}')\}^{P}
\Delta^{*}({\bf r}_{1},{\bf r}_{o}),
\end{equation}
and
\begin{equation}
\Delta^{*}(m) = T \sum_{n'}
\sum_{m'}V_{mm'}
{\Delta^{*}(m')\over \omega_{n'}^{2}
+\epsilon_{m'}^{2}},
\end{equation}
where
\begin{equation}
 V({\bf r}-{\bf r}')=\sum_{\vec q}\theta(\omega_{D}-\omega_{\vec q})
e^{i{\vec q}\cdot ({\bf r}-{\bf r}')},
\end{equation}
and 
\begin{equation}
V_{mm'} = \gamma^{2}\int\int 
 V({\bf r}-{\bf r}')
\psi_{m}^{*}({\bf r})\psi^{*}_{\overline m}({\bf r}')
\psi_{m'}({\bf r}) \psi_{\overline m'}({\bf r}')
d{\bf r}d{\bf r}'.
\end{equation}
In this case, the effect of impurities on $T_{c}$ is more
complicated because of the nonlocal nature of the
pairing interaction. Nevertheless the result may not be much different from
that obtained from an Einstein model.

\section {Conclusion}

Using the Eliashberg formalism, we reconsidered the impurity effect on the 
transition temperature of superconductors.
It is shown that the correspondence principle, which relates strong-coupling 
and weak-coupling theories, works only when Anderson's pairing condition is 
used.
The change of the electron density of states caused by the impurity scattering
may be negligible in practice. Whereas the phonon-mediated interaction 
decreases by the same weak localization correction term as that of the
conductivity.
Consequently, for strongly localized states the phonon-mediated interaction 
is exponentially small.

\vspace{2pc}
\centerline{\bf Acknowledgments}

This work has been supported by the Brainpool project of KOSEF and the MOST.
I am grateful to Prof. YunKyu Bang for discussions.

\vfill\eject
{\bf Table I.} Mean free path and phonon-mediated interaction
in dirty, weak localization and strong localization limits.
Here $\ell$ and $L$ are the elastic and inelastic mean free paths and 
$\alpha$ denotes the inverse localization length.

\vspace{2pc}

\begin{tabular}{lrrr}\hline
{disorder limit } & \hspace{2pc} { dirty } & \hspace{3pc} { weak localization  \hspace{2pc}} &\hspace{1pc}  { strong localization }  \\ \hline
$\hspace{2pc}\ell$  & $ \sim 100\AA$ & $\sim 10\AA\hspace{4pc}$
                   & $\sim 1\AA$ \hspace{3.5pc} \vspace{1pc}\\ 
$\hspace{1pc}V_{mm'}$  & $V$\hspace{1.0pc} & $V[1-{2\over\pi k_{F}\ell}ln(L/\ell)] \hspace{1.0pc}(2d)$
   & $\sim exp(-\alpha L)$\hspace{2pc} \\ 
  &   &  $V[1-{3\over(k_{F}\ell)^{2}}(1-{\ell\over L})] \hspace{1.0pc}(3d)$
                           & \\ \hline
\end{tabular}

\end{document}